# Optimization of gain uniformity in thermal bonding Micromegas for the PandaX-III experiment

Yunzhi Peng [a,b], Yuanchun Liu [c], Zhiyong Zhang [a,b], Shaobo Wang [c], Jianbei Liu [a,b], Ming Shao [a,b], Yi Zhou [a,b]

[a] State Key Laboratory of Particle Detection and Electronics, University of Science and Technology of China, Hefei, 230026, China

[b] Department of Modern Physics, University of Science and Technology of China, Hefei, 230026, China

[c] INPAC and School of Physics and Astronomy, Shanghai Jiao Tong University, MOE Key Lab for Particle Physics, Astrophysics and Cosmology, Shanghai Key Laboratory for Particle Physics and Cosmology, Shanghai 200240, China

**Abstract**

Micro-pattern gas detectors (MPGDs) are widely utilized in physics experiments owing to their excellent spatial resolution and high-rate capabilities. Within the PandaX-III experiment, which aims to investigate neutrinoless double beta decay, Micromegas detectors serve as charge readout devices. High energy resolution is a critical requirement for the readout plane in this context, and gain uniformity significantly impacts the achievable resolution, primarily because of the extended tracks of primary ionization electrons.

However, scaling up MPGDs to larger active areas exacerbates the challenge of maintaining gain uniformity, and effectively controlling the uniformity of the avalanche gap is a key factor in the detector manufacturing process via the thermal bonding method. This study demonstrates that optimizing the thermal bonding films specifically at the detector edges can effectively improve the gain uniformity, achieving a gain uniformity of < 5% over the entire 200×200 mm² active area in a 1 bar Ar/isobutane (96.5/3.5) gas mixture. Additionally, the gain uniformity of approximately 14% was characterized at high pressures of up to 10 bar, revealing promising potential for high resolution measurements in the PandaX-III experiment and other high-pressure applications.

Keywords: gain uniformity, Micromegas detector, thermal bonding method, PandaX-III experiment,

## 1.Introduction

Micro-pattern gas detectors (MPGDs) are extensively employed in physics experiments (e.g., ATLAS, COMPASS and PandaX-III), neutron imaging, and medical diagnostics [1,2] owing to their outstanding performance characteristics, including high spatial resolution, high-rate capability, and radiation hardness [3,4].

The PandaX-III experiment, which aims to detect neutrinoless double beta decay (NLDBD) at the China Jinping Underground Laboratory, employs a 10 bar $^{136}$Xe Time Projection Chamber (TPC) equipped with a readout plane comprising 52 Micromegas detectors, each measuring 200×200 mm² [5]. This setup is designed to achieve high granularity, long-term stability, good energy resolution, and a low background radioactivity.

In this context, gain uniformity is a crucial determinant of energy resolution. The intrinsic energy resolution of the Micromegas detector, considering only avalanche fluctuations, is approximately 0.7% at the 2.458 MeV Q-value. However, the extended electron tracks characteristic (~200 mm) of NLDBD events experience spatially varying gains across the detector active area, which further degrades the achievable energy resolution.

While uniformity correction techniques can partially mitigate this effect, the X-Y strip readout currently employed cannot accurately reconstruct all individual hit positions, thus preventing complete correction of the non-uniformity. In contrast, a pixel readout scheme enables position reconstruction at the pixel scale, allowing nearly perfect correction of the gain non-uniformity effect and consequently contributing to improved energy resolution.

Preliminary simulations confirm that after uniformity correction, an X-Y strip readout Micromegas with 5% gain uniformity at 10 bar achieves the target energy resolution of 3% at 2.458 MeV[5]. Under identical conditions, a pixel readout system reaches about 1.2% energy resolution despite 10-20% gain uniformity. Therefore, a Micromegas detector with this uniformity level at 10 bar is suitable for future pixel readout systems.

This study presents an optimization strategy to improve the gain uniformity of thermal bonding Micromegas detectors for PandaX-III experiment. The paper is structured as follows: Section 2 introduces the thermal bonding Micromegas detectors utilized in PandaX - III experiment. Section 3 analyzes the avalanche gap difference between the edge and central regions through simulations and experiments; Section 4 details the optimization method and the improved results; and Sections 5 and 6 discuss high-pressure performance and future improvements.

## 2. The characteristics of thermal bonding Micromegas detectors for PandaX-III experiment

### 2.1 Manufacturing processes

The thermal bonding method[6] developed in the University of Science and Technology of China (USTC) provides a cost-effective, etching-free solution for manufacturing Micromegas detectors.). **Figure 1** shows the photograph, top-view and cross-sectional schematics of the thermal bonding Micromegas detector used in PandaX-III experiment. Edge films and spacer pillars (both shown in gray) are made of the same thermal bonding film material. They only exhibit distinct geometries and functions: the edge films form a 3.5 mm-wide frame that provides mechanical anchoring and uniform tension for the mesh, and the spacer pillars are 1 mm-diameter cylinders designed to maintain a nominal 100 μm avalanche gap while mitigating deformation. The cross-sectional view (**Figure 1**, bottom) shows the layered architecture: from top to bottom, the black dashed line represents the micro-mesh, the gray components are the edge films and spacer pillars, the blue and yellow layers are flexible PCB substrates with integrated copper readout strips, and the orange base layer represents the reinforced base plate that ensures mechanical rigidity and serves as a planar reference.

The thermal bonding film comprises three functional layers, as illustrated in **Figure 2**. The central layer is a 100-μm-thick insulating dielectric, typically polyethylene terephthalate (PET) for mechanical support. The upper and lower layers are thermal bonding adhesive with thicknesses of 25 μm each, which melts at high-temperature processing and re-solidifies upon cooling.

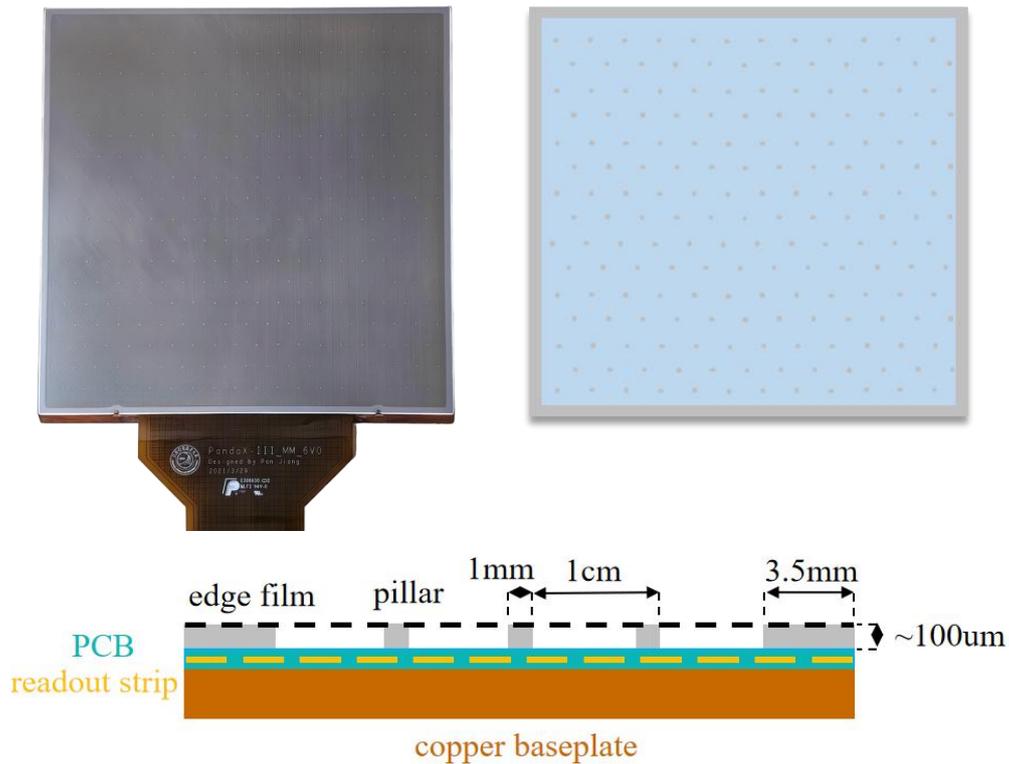

**Figure 1.** The photograph (top-left), top-view (top-right) and cross-sectional (bottom) schematics of the thermal bonding Micromegas used in PandaX-III experiment.

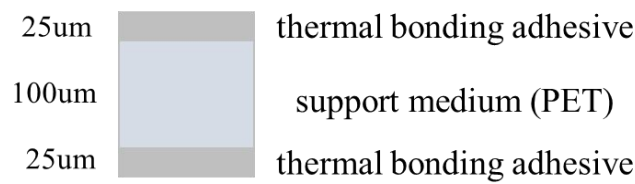

**Figure 2.** Schematic illustration of the thermal bonding film.

## 2.2 Gain uniformity

To characterize the gain uniformity, a $^{55}$Fe source is employed to irradiate the detector, **Figure 3** illustrates a schematic and photograph of the experimental setup. The uncollimated X-ray source is positioned 30 cm above the detector for full active-area coverage. The detector signals are passed through an interface board and connected to the electronics system, processed by the front-end electronics designed by USTC[7], and subsequently sent to a data acquisition system for uniformity analysis.

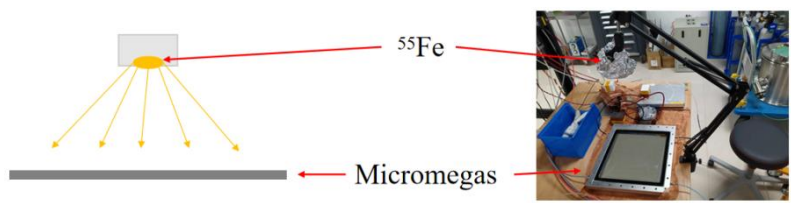

**Figure 3**. Schematic diagram (left) and photograph (right) of the experimental setup

**Figure 4** presents a representative gain uniformity measurement of a thermal bonding Micromegas detector operating in an Ar/isobutane (96.5/3.5) gas mixture at 1 bar using flowing gas. The ADC values of the X-ray energy peaks are used to represent the gain. The left picture displays a two-dimensional gain distribution across the 192 mm × 192 mm active area, segmented into a 32 × 32 grid (1024 cells), where color intensity represents local ADC values. The right picture shows the histogram of all the ADC values. A significant gain reduction at the periphery compared with the central region is observed.

Gain uniformity is quantified as the ratio of the standard deviation of gain values to the mean gain within a defined region. The central region exhibits 5% gain uniformity (derived from the Gaussian fit in **Figure 4**, right), whereas the inclusion of peripheral regions decreases the gain uniformity above 10%. Consequently, compensating for the gain reduction at the periphery to match the values in the central region is a crucial optimization strategy. Moreover, the origin of the gain non-uniformity, especially the discrepancy between the central region and the edge, will be analyzed hereinafter.

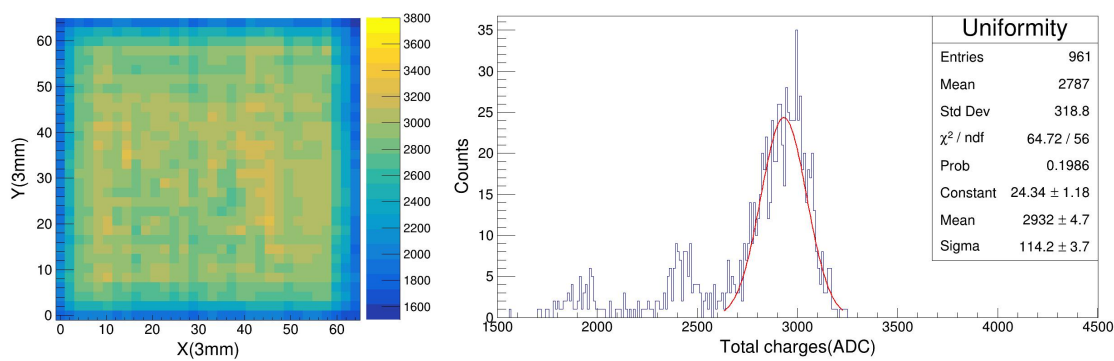

**Figure 4**. Two-dimensional distribution (left) and one-dimensional histogram (right) of gain uniformity in a typical thermal bonding Micromegas detector.

# 3. Main factors affecting gain uniformity

## 3.1 Simulation study

The abovementioned differences in gain between the edge and central regions are likely attributed to variations in the thickness of the avalanche gap.

$$G = e^{AV+Bd} \qquad (1)$$

Here, $G$ represents the gain magnitude, whereas $A$ and $B$ are coefficients that include factors such as gas composition, gas proportion, pressure, and electric field intensity. $V$ denotes the potential difference between the mesh and the anode, and $d$ represents the distance between the mesh and the anode, which is also the avalanche gap size. At a given instant when the Micromegas detector is operational, $A$, $B$, and $V$ are constant values, whereas $d$ usually varies with position. This variation in $d$ leads to changes in the gain $G$.

**Figure 5** (left) shows the calculated gain as a function of avalanche gap thickness (10–500 μm) in an Ar/CO (93/7) mixture at 1 bar, obtained using the Magboltz-based gain formula in Garfield++ [6]. The gain exhibits significant variation across a broad range of avalanche gap thicknesses. **Figure 5** (right) displays the simulated gain against gap thickness in an Ar/isobutane (96.5/3.5) mixture at 1 bar under a mesh voltage of –400 V, as computed with the electron avalanche simulation module in Garfield++. A strong inverse correlation is observed in the 100-140 μm range, where the gain exhibits an exponential decrease of about 25% per 10 μm increment in gap thickness. This pronounced dependence demonstrates the critical sensitivity of electron multiplication processes to gap thickness variations in the avalanche region.

Such a relationship explains the peripheral gain reduction observed in **Figure 4** relative to the central active region. The geometric non-uniformity in practical detectors results in localized gap expansion at edges, reducing the electric field strength. The diminished field strength directly compromises the electron multiplication efficiency, resulting in localized gain suppression.

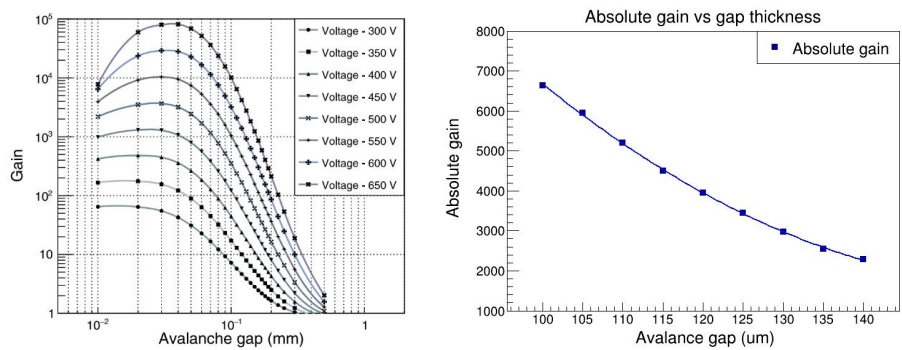

**Figure 5.** Gain distribution variation with avalanche gap sizes by formula calculations (left) [6] and simulations (right).

## 3.2 Experimental test of avalanche gaps and gains

To validate the above simulation results, precise measurements of the avalanche gap thickness and the corresponding gain were conducted. The measurement procedure specifically compared regions displaying different gain characteristics.

A Leica digital confocal microscope (Model DM2700M) was used to measure the gap thickness. **Figure 6** presents the experimental setup, including both the testing environment and the microscope configuration. Measurements were conducted at six locations: adjacent to the edge film and at distances corresponding to 1, 2, 3, 5, and 7 spacer pillars from the edge. The spatial relationships are detailed in **Figure 7**, where the blue line indicates the edge film, blue dots represent spacer pillars, and red dots mark measurement positions.

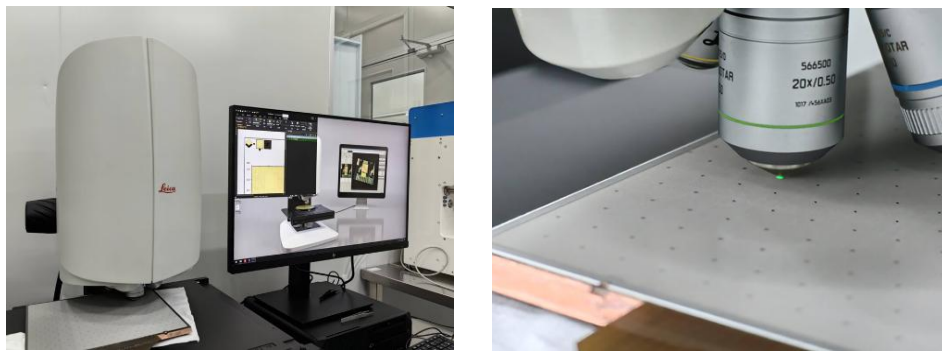

**Figure 6**. Photograph of the Leica digital confocal microscope device (left) and a close-up shot of the measurement of the avalanche gap thickness (right).

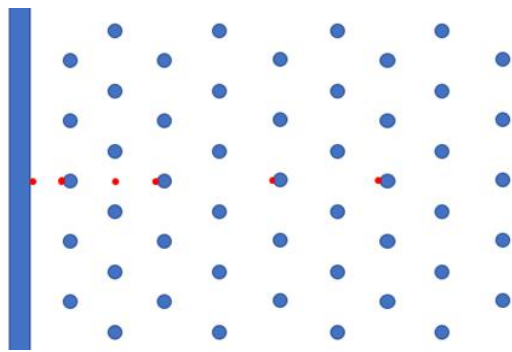

**Figure 7**. Schematic layout of the six measurement points.

The reconstructed 3D topography in **Figure 8** employs color-mapped height variation, where the pink and green surfaces correspond to the micro-mesh upper plane and PCB substrate, respectively. The thickness of woven micro-mesh structure increases at wire intersections. Optical occlusion by the mesh causes reconstruction artifacts connecting the mesh underside to the PCB, despite the physical suspension of the mesh.

The height profile along the transverse direction shown in **Figure 9** represents the measured elevation (blue trace), with the PCB substrate forming the lower baseline and the micro-mesh exhibiting upper peaks. Yellow markers define regions for spatial averaging of height measurements. The height differential is derived by calculating the height difference between the averaged micro-mesh surface and PCB reference plane within adjacent marker zones. This computation was performed at eight equally spaced transverse positions presented in **Figure 9**, with the arithmetic mean yielding the final height differential illustrated in **Table 1**. Using the known 24 μm micro-wire diameter, the avalanche gap in **Table 1** is derived by subtracting the wire diameter from the measured height differential.

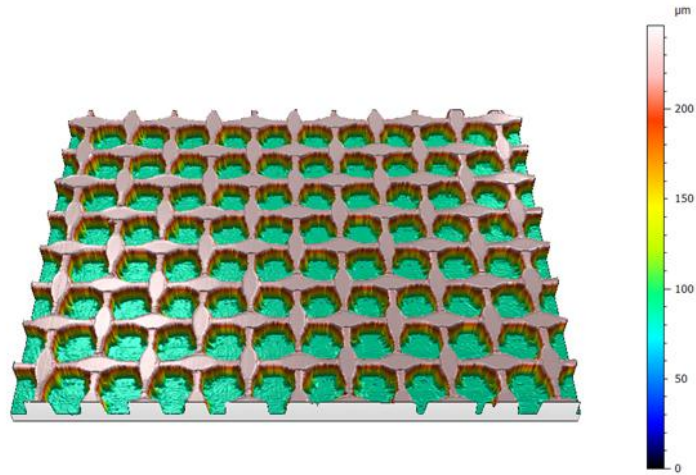

**Figure 8**. Reconstructed 3D topography at one measurement point

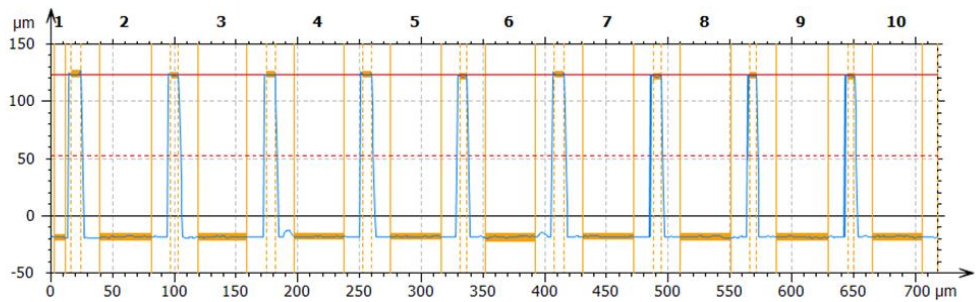

**Figure 9**. Cross-sectional profile and height differential measurement

**Table 1**. The height differential between mesh's upper face and PCB, and the avalanche gap at 6 measurement points.

| No. of measurement points | 1 | 2 | 3 | 4 | 5 | 6 |
|---|---|---|---|---|---|---|
| Height differential (um) | 152.8 | 136.6 | 135.1 | 133.5 | 131.6 | 129.3 |
| Avalanche gap (um) | 128.8 | 112.6 | 111.1 | 109.5 | 107.6 | 105.3 |

Following the gap thickness measurements, the same Micromegas detector was employed for gain characterization at identical spatial positions within an Ar/isobutane (96.5/3.5) flow chamber, as illustrated in **Figure 10**. An $^{55}$Fe source was placed at the six measurement points predefined in **Figure 7**, with the gain recorded at a mesh voltage of -400 V. The measurement method is analogous to that depicted in **Figure 3**. The most significant disparity lies in the fact that only the gains of a limited number of points were measured using a collimated source, rather than the entire active area with an uncollimated source.

Both simulated and experimental gain values, normalized to the data in 100 μm gap, are compared in **Figure 11**. Positioning uncertainty introduces transverse error bars from gap thickness variations.

Key observations include a gain reduction near the edge film correlated with increasing gap dimension, and stable high gain maintained in the central active region. The experimental results are in good agreement with simulation, confirming edge-induced gap deformation as the governing mechanism. Moreover, significant variations in gap thickness and gain exist even at a transverse distance of 1 mm in edge areas. Consequently, homogenizing the gap thickness between central and peripheral regions would further improve local energy resolution in edge areas.

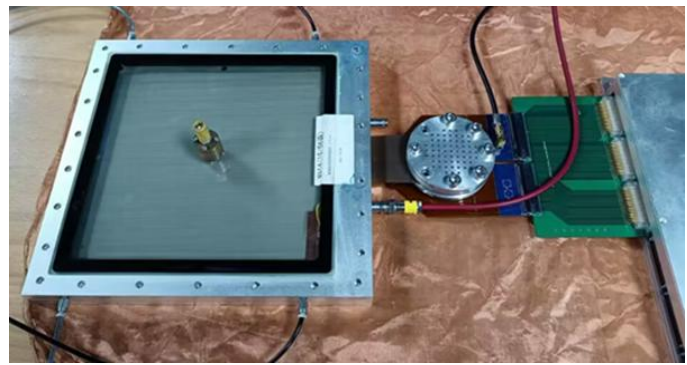

**Figure 10**. Picture of the gain test at six points.

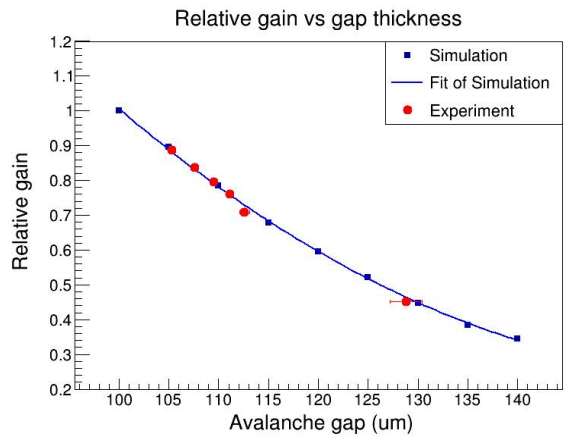

**Figure 11**. Relationship between avalanche gap thickness and gain from simulation and experiment.

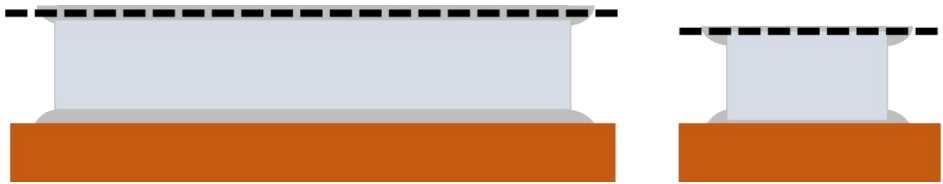

**Figure 12.** Schematic diagram of the possible cross-sections of the edge film (left) and central region spacers (right) of thermal bonding Micromegas detectors.

Following the aforementioned simulation and experiment, the potential structures of edge films and central spacer pillars after thermal bonding were put forward. The cross-sectional geometries of them are compared in **Figure 12** (left and right, respectively), highlighting their distinct effects on the local avalanche gap dimensions.

Structural differences arise from continuous edge film distribution versus discrete pillar arrangement. During thermal bonding, spacer pillars exhibit omnidirectional adhesive extrusion due to axisymmetric confinement, resulting in a final thickness determined primarily by the 100 μm support medium. Conversely, edge films demonstrate constrained adhesive flow: the upper adhesive layers adjacent to the mesh extrude upward and laterally, while the lower layers bonded to the PCB are limited to lateral flow.

Geometric factors further differentiate their behavior: the 3.5 mm-wide continuous edge film possesses a higher adhesive volume-to-area ratio than the 1 mm-diameter discrete pillars, promoting incomplete adhesive expulsion during compression. Analysis of the post-bonding thickness indicates that the pillar regions maintain a thickness of approximately 100 μm, whereas edge films retain significant residual adhesive above the support layer.

Consequently, the peripheral regions develop larger avalanche gaps than central zones, reducing electric field strength and degrading electron multiplication efficiency near detector edges. This geometric non-uniformity induced by the fabrication process represents the key mechanism underlying the gain differences between edge and central regions.

## 4. Optimization methods and performance improvements

Based on the identified mechanism of thickness non-uniform, achieving a post-bonding edge film thickness of approximately 100 μm represents a critical optimization target. Three methods were experimentally investigated: thermal bonding parameter optimization (temperature, pressure, duration), laser perforation of edge films, and structural redesign of multilayer edge films.

### 4.1 Optimization of thermal bonding parameters

Optimizing thermal bonding parameters effectively reduces the post-bonding thickness. Significant improvements were observed with initial increases in temperature, pressure, and duration. However, beyond critical thresholds (temperature > 210 °C, pressure > 1.8 MPa, duration > 90 s), the process exhibited diminishing returns. The edge film thickness stabilized at a level still exceeding the pillar height, indicating the need for alternative approaches.

## 4.2 Edge film optimization via laser-drilled micro-perforations

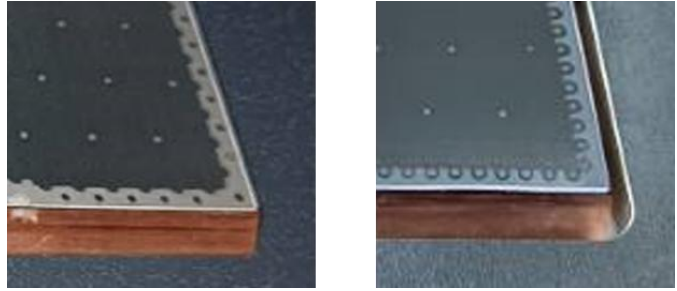

Figure 13. Two proposed perforation schemes for edge films.

Two laser perforation schemes were applied to the edge films in detector prototypes presented in **Figure 13**. The introduced 500 μm diameter pores promoted adhesive flow and extrusion during thermal bonding, effectively reducing thickness. While achieving outstanding gain uniformity of approximately 5%, this approach introduced critical limitations: increased fabrication complexity, a high sparking propensity, and operational instability at 10 bar gas pressure. Consequently, this method was discontinued.

## 4.3 Edge film optimization via multilayer thickness tuning

To resolve persistent post-bonding thickness discrepancies, a dual-material strategy was developed: engineered adhesive films with customized compressibility were used for edge regions, while standard 25/100/25 μm thermal bonding films were retained for pillars. This approach enables precise thickness matching between peripheral and central regions.

Two thickness optimization routes for new edge films were explored: one using thicker adhesive layers with reduced support medium dimensions, and another maintaining conventional adhesive thickness with minimized support geometry.

Limited commercial availability of suitable films necessitated the development of custom composite materials, fabricated via precision lamination of separately sourced adhesive and support layers to achieve micrometer-scale thickness control. Owing to the stochastic nature of adhesive flow during thermal bonding, extensive experimental validation was conducted. Multiple detector prototypes were produced by edge films with systematically varied adhesive–support–adhesive layer ratios shown in **Figure 14**, each evaluated through rigorous gain uniformity measurements.

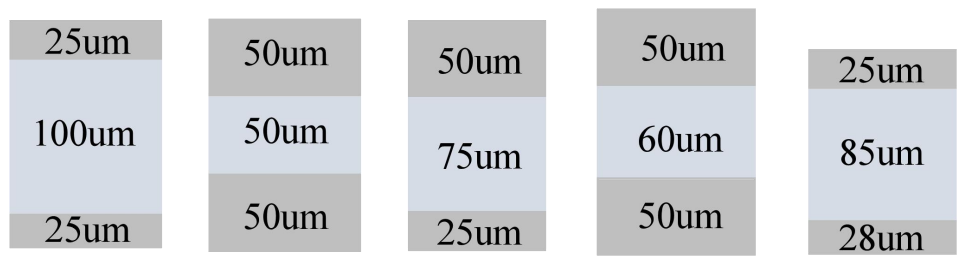

Figure 14. Cross-sectional schematic of several thermal bonding film architectures

(light blue: supporting medium; gray: thermal bonding adhesive layers).

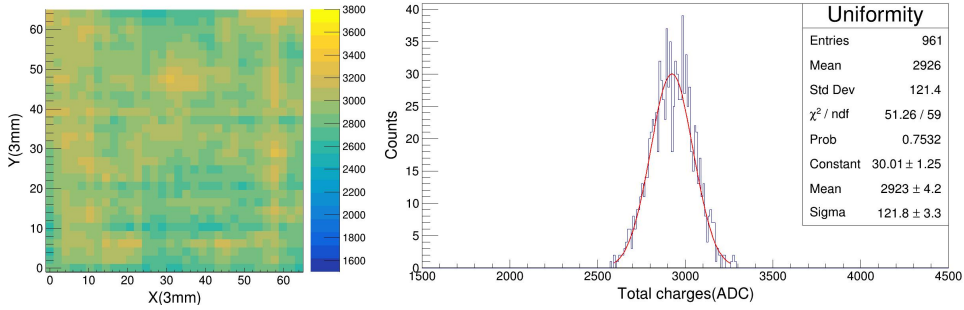

**Figure 15**. Two-dimensional distribution (left) and one-dimensional histogram (right) of gain in a thermal bonding Micromegas with 25/85/25 μm edge films.

The 25/85/25 μm edge film configuration demonstrated optimal performance, as shown in **Figure 15**. The detector exhibited excellent gain uniformity, with a relative standard deviation of 4.5% over the active area and a central region Gaussian dispersion (σ/mean) of 4.2%.

These results validate the optimized multilayer edge film approach as an effective solution for reducing the gap non-uniformity in thermal bonding Micromegas. The methodology establishes a reproducible fabrication framework suitable for next-generation particle detectors requiring high spatial homogeneity and scalable production.

## 5. Gain uniformity under high pressure

All gain uniformity results reported above were obtained at normal pressure using flowing gas mixtures. However, the PandaX-III experiment requires detector operation at 10 bar. Consequently, a gain uniformity test under different pressures was carried out.

A dedicated high-pressure vessel named miniTPC, capable of operating at up to 17 bar and equipped with a field cage defining a 20 cm drift region, was used to house the Micromegas detector (**Figure 16**). The vessel was filled with Ar/isobutane (97.5/2.5) mixture at pressures of 1, 3, 5, 7, and 10 bar[8]. $^{37}$Ar was used as a spatially uniform calibration source due to its excellent diffusion properties.

Similar to the above flowing-gas tests, the active area of Micromegas was divided into a 21×21 grid (441 small cells). However, gain calculations in peripheral regions were unreliable due to electric field distortion and the significant gain variations inherent to the non-optimized Micromegas detector used in the test. Therefore, uniformity was evaluated using the central 19×19 region (361 cells).

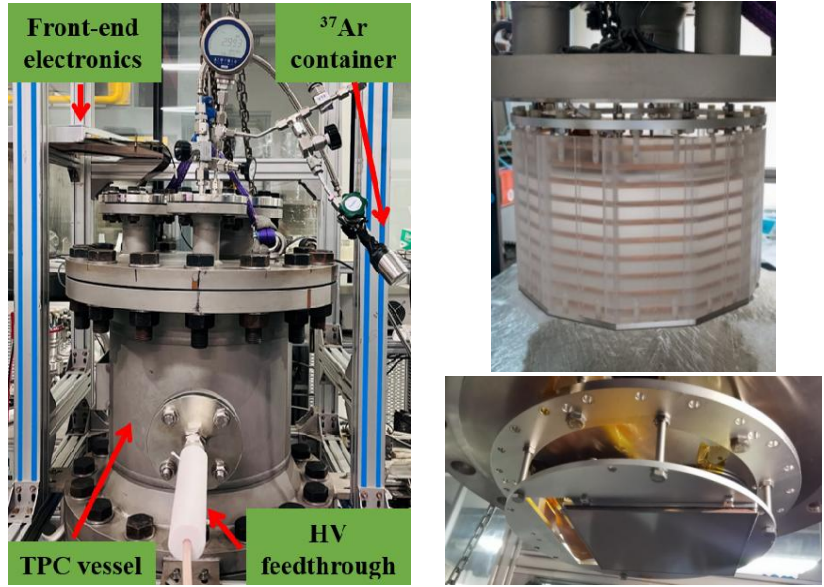

**Figure 16** Pictures of the miniTPC[8] (left), field cage(top-right) and the Micromegas(bottom-right) installed in it.

The experimental results are presented in **Figure 17.** Under a pressure of 1 bar, the observed uniformity is approximately 5% when the gain ranges from 1000-4000, which is similar to the results obtained from the gas mixture of Ar/isobutane (96.5/3.5). It confirms the insensitivity of uniformity to this minor gas composition change. A clear degradation in uniformity was observed with increasing pressure. At a fixed gain of approximately 1000, the gain non-uniformity at 10 bar exhibited an increase by a factor of 2–3 in comparison to that at 1 bar, reaching approximately 14%.

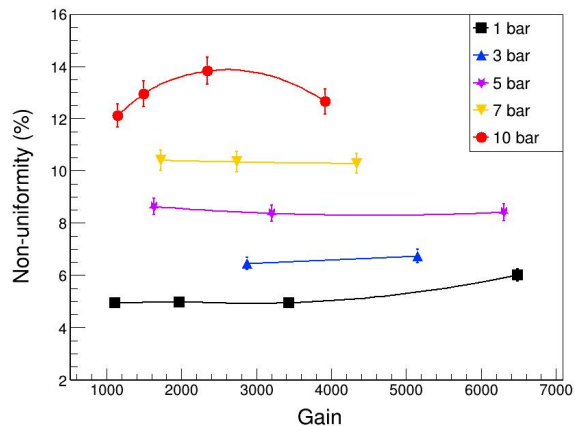

**Figure 17** Correlations between non-uniformity and gains at different gas pressures.

## 6. Future directions for optimization

Further improvement in gain uniformity can be pursued through PCB optimization, as the ~14% uniformity achieved at 10 bar is largely limited by PCB surface height variations. While using a flatter PCB helps reduce these variations, further uniformity enhancement alone represents a diminishing-returns approach for improving the energy resolution in PandaX-III.

As noted in the introduction, preliminary simulations confirm that with uniformity correction applied, a pixel readout system achieves an energy resolution of 1.2% at 2.458 MeV despite 14% gain non-uniformity under 10 bar operating pressure. These findings validate that developing a pixel readout Micromegas — building upon the established uniformity optimization framework—represents a more promising approach for attaining superior energy resolution, warranting further investigation in future work.

## 7. Conclusion and outlook

This work systematically addresses the gain non-uniformity between edge and central regions in thermal bonding Micromegas detectors, tracing its origin to the geometric non-uniformity of the avalanche gap. Combined experimental and simulation analyses demonstrate that thickness discrepancies between edge films and central pillars reduce electric field strength in peripheral regions, thereby suppressing electron multiplication efficiency.

Among three solutions investigated, the optimized multilayer edge film strategy proved most effective. The custom 25/85/25 μm configuration achieved a gain uniformity of 4.5% (relative standard deviation) across the active area and 4.2% (σ/mean) in the central region at 1 bar, maintaining approximately 14% uniformity at 10 bar. This approach establishes a reproducible fabrication framework for large-area MPGDs requiring high spatial homogeneity. In the future, the pixel readout Micromegas integrated with uniformity optimization will be explored to meet the energy resolution requirement of the PandaX - III experiment.

## Acknowledgments

The authors acknowledge significant technical contributions from Dr. Sicheng Wen and the team at JIANWEI Scientific Instruments Technology Co., Ltd for their expertise in detector fabrication. And the setup and data acquisition efforts of Dr. Wenming Zhang in the $^{37}$Ar test, are also highly appreciated. Special recognition is extended to the Center for Micro and Nanoscale Research and Fabrication at the University of Science and Technology of China (USTC) for providing access to advanced characterization facilities, and we particularly thank Mr. Haitao Liu for his great help with the digital confocal microscope.

## Funding

This work was supported by the Program of National Natural Science Foundation of China [Grant numbers 12125505 and 12522514] and the Double First-Class University Project Foundation of USTC.